# XAS study of the local environment of impurities in doped $TiO_2$ thin films.

C.E. Rodríguez Torres[1], A. F. Cabrera[1], L.A. Errico[1], S. Duhalde[2], M. Rentería[1], F. Golmar[2] and F.H. Sánchez[1]

[1] *Departamento de Física e IFLP (CONICET), Facultad de Ciencias Exactas, Universidad Nacional de La Plata*

[2] *Lab. de Ablación Láser, FI-UBA*

**Abstract**

In this work we present an X-ray Absorption Spectroscopy characterization of the local environment of the impurity in room temperature ferromagnetic anatase $TiO_2$ thin films doped with Co, Ni, Cu, or Zn, deposited on $LaAlO_3$ substrate by Pulsed Laser Deposition. It was found that there is a considerable amount of impurity atoms substituting Ti in $TiO_2$ anatase, although the presence of metal transition monoxide clusters can not be discarded. From our results we infer that the observed room temperature ferromagnetism of the samples could be assigned to the metal transition atoms replacing Ti in $TiO_2$ anatase.

Diluted magnetic semiconductors, Thin films, EXAFS, XANES

## 1. Introduction

The discovery of room-temperature ferromagnetism (RTF) in Co-doped anatase [1] and rutile [2] $TiO_2$ has stimulated intense theoretical and experimental studies of these systems due to their potential applications in spintronic technology. In effect, dilute magnetic semiconductors (DMS), produced by doping non magnetic semiconductors with transition metals, combine their electric conductivity with ferromagnetism and optical transparency, thereby opening up the possibility of new device concepts. But, the site of the impurity is still under debate and while some authors claimed that the presence of ferromagnetic metallic clusters can not be completely excluded, others have concluded that ferromagnetism is a consequence of atomic scale doping [3].

Recently, we reported the experimental observation of unexpected and significant RTF in Cu-doped $TiO_2$ films [4], equivalent to what would be expected if, on the average, each Cu atom bears a magnetic moment of about 1.5 $\mu_B$. This result indicates that the presence of impurity clustering is not necessary to obtain ferromagnetic order. A large magnetic moment was also obtained from *ab initio* calculations, but only if an oxygen vacancy in the nearest-neighbor shell of Cu is present [4].

In order to clarify the role of impurities in DMS on the presence of magnetism it is crucial to determine the local environment of magnetic and non magnetic impurities. In this work we present an EXAFS (Extended X-ray Absorption Fine Structure) and XANES (X-ray Absorption Near Edge Spectroscopy) characterization of the local environment of transition metal (TM)-doped anatase $TiO_2$ thin films (TM=Co, Ni, Cu, and Zn).

**2. Experimental**

Thin films of 10 at.% TM (Co, Ni, or Cu) and 5 at.% Zn-doped $TiO_2$ were deposited on $LaAlO_3$ (001) substrate by Pulsed Laser Deposition (PLD) using a Nd:YAG laser operating at 266 nm. The substrate temperature, laser energy density, oxygen pressure, and pulse repetition rate were 800°C, 2 J/cm$^2$, 20 Pa, and 10 Hz, respectively.

X-ray absorption spectroscopy (XAS) measurements were taken at room temperature in fluorescence mode at the TM K-edge, using a Si (111) monochromator at the XAS beamline of LNLS (Campinas, Brasil).

**3. Results and discussions**

The structural and magnetic characterization of all these samples were already published elsewhere [4,5]. All films are RT ferromagnetic, transparent to visible light and strongly textured. X-ray diffraction (XRD) studies (not presented here) showed only the (001) reflection of the anatase structure for pure and TM-doped $TiO_2$ films.

Fig 1 shows XANES spectra of the doped films. They are compared with those of the TM oxide references that present more similarities with the former ones. The agreement with these references (all monoxides) is excellent in the case of Cu, rather good in the Ni and Co ones, and poor in the case of Zn. These results indicate that the TMs in the films are in the 2+ formal oxidation state. These spectra are also very different from those of the metallic elements, allowing us to rule out the presence of TM precipitates.

In Fig. 2 the EXAFS results are summarized. The oscillations observed in the films and their corresponding Fourier transforms (FT) are compared with those of the same oxide references used in Fig. 1. In the case of the Co-doped film, major similarities between the films oscillations and any of the measured references (CoO, $Co_2O_3$, and metallic Co) ones were not found. It may exist some correspondence with CoO. In the corresponding FT some coincidences with $Co_2O_3$, not shown here, and with the CoO spectra were found (especially in peak positions of the first two layers). Also, the Co-environment, reflected in both oscillation and FT, is very similar to that of $CoTiO_3$, in agreement with the results reported in Ref. [6].

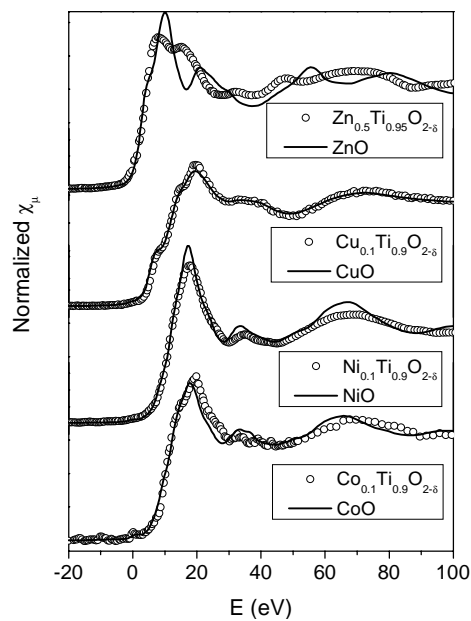

Figure 1: TM (TM= Co, Ni and Cu) K-edge XANES spectra of $TM_{0.1}Ti_{0.9}O_{2-\delta}$ and $Zn_{0.05}Ti_{0.95}O_{2-\delta}$ thin films on $LaAlO_3$ (001). The results obtained for TM monoxide reference samples are also shown for comparison.

The oscillations and the corresponding FT of Ni and Cu-doped films have similarities with those of NiO and CuO, respectively. In the case of Cu-doped one, the agreement is better. In the FT of the Ni-doped film the attenuation experienced by peaks corresponding to the second and next neighbour layers is evident. No similarities between the Zn-doped film and ZnO EXAFS oscillations were found. However, in the FT there are some coincidences, especially in the first coordination layer and in the position of the second one.

Then, from the comparison of XANES and EXAFS spectra of the doped films with those of the references, there are some coincidences with the corresponding TM monoxides. The agreement is excellent in the case of Cu, rather good for Ni and poor in the cases of Co and Zn (worst for the Zn one). However, in all cases there is a high degree of similarity between the FT of the doped films and those corresponding to their TM monoxides, especially in the first coordination layer.

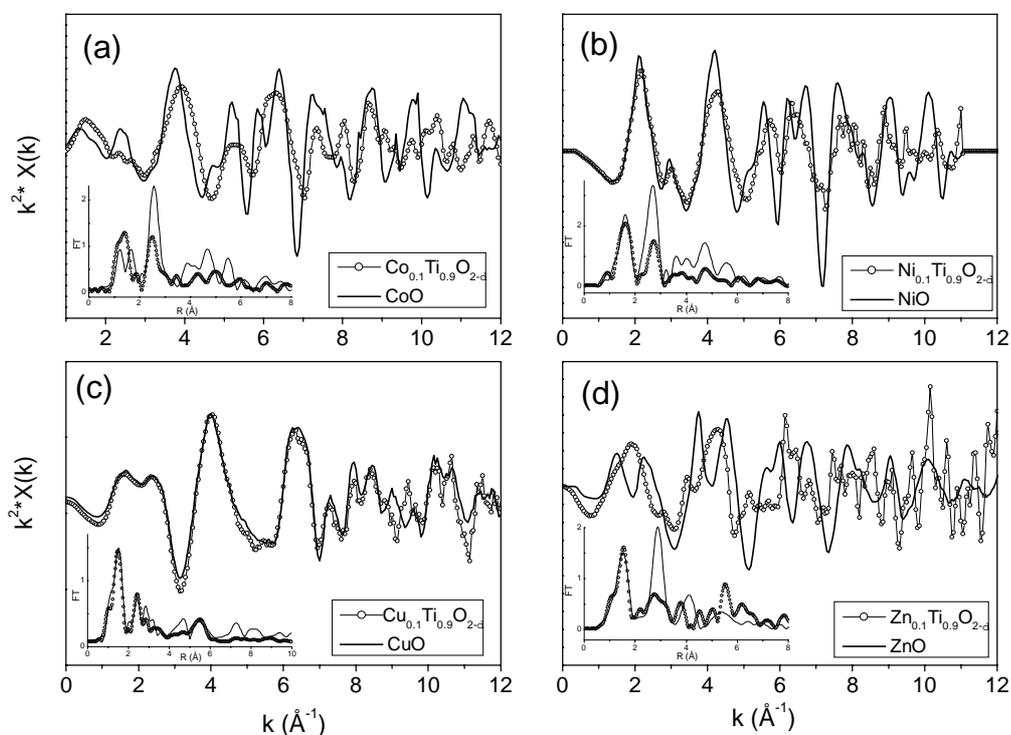

Figure 2: EXAFS oscillations and their corresponding Fourier transforms (inset) of: (a) Co-doped, (b) Ni-doped, (c) Cu-doped, and (d) Zn-doped $TiO_2$ thin films. In all cases the results for the respective TM monoxide are also shown for comparison. The FTs were performed in the range k=1.5-12 Å$^{-1}$ with $k^2$ weighting in the oscillation.

In Fig. 3 it can be seen that there is also an strong similarity between the FT of Ti in $TiO_2$ anatase and those of TM in the films, in the fact that they present a second coordination peak smaller than the first one that corresponds to oxygen coordination. While CoO and NiO present the NaCl structure, CuO has the tenorite and ZnO the wurzite ones. All these oxides have oxygen nearest neighbours (NN) at similar distances to those of Ti-O in $TiO_2$ anatase (see Table I). While the first two oxides have octahedral coordination (regular six-fold coordination for the cation) as Ti in $TiO_2$, the other two have their cations four-fold coordinated. In the case of CuO the 4 ONN are in the plane that contains the central Cu, as it is the case of the 4 ONN around Ti in $TiO_2$ anatase. In ZnO the cation has a tetrahedral coordination with its ONN. Then it is expected that the first FT peak of a TM atom substituting Ti in $TiO_2$ anatase be similar to the corresponding peak of the TM oxide. In the second coordination sphere of CoO and NiO, the TM has 12 TM NN at 2.95Å (Co) and 3.02 Å (Ni). In CuO, Cu has 4 Cu NN at 2.90 Å and 4 at 3.08 Å. In ZnO, Zn has 12 Zn NN at a mean distance of 3.23 Å. These distances (not the coordination numbers) are comparable with the Ti-Ti ones (4 Ti NN at 3.04 Å) in anatase. Then, it is presumed that the second FT peak of TM in anatase could be similar to that of TM in its oxide in the case of Cu and smaller in intensity but similar in position in the cases of Co, Ni and Zn. In our films the FT second coordination sphere peak coincides in position with the TM oxide one but with small intensity, suggesting that considerable amount of TM is substituting Ti in $TiO_2$ anatase, although the presence of TM oxide clusters can not be discarded. Since the TM oxides (at least in bulk) are paramagnetic at RT, from these results it is possible to ascribe the ferromagnetism found in these samples to the presence of TM substituting Ti in TiO$_2$ anatase, probably with the presence of one or two NN oxygen vacancies, in agreement with previous *ab initio* predictions [7].

|  | CoO |  | NiO |  | CuO |  | ZnO |  | TiO$_2$ anatase |  |
|---|---|---|---|---|---|---|---|---|---|---|
|  | N/A | R (Å) | N/A | R (Å) | N/A | R (Å) | N/A | R (Å) | N/A | R (Å) |
| 1$^{st}$ layer | 6 O | 2.08 | 6 O | 2.13 | 4 O | 1.95 | 3 O<br>1 O | 1.97<br>1.99 | 4 O<br>2 O | 1.94<br>1.96 |
| 2$^{nd}$ layer | 12 Co | 2.95 | 12 Ni | 3.02 | 2 O<br>4 Cu<br>4 Cu | 2.78<br>2.90<br>3.08 | 6 Zn<br>1 O<br>6 Zn | 3.21<br>3.22<br>3.25 | 4 Ti | 3.04 |

Table I: 1$^{st}$ and 2$^{nd}$ shells of neighboring atoms around the cation in the oxides quoted. N is the number of A atoms located at a distance R from the cation.

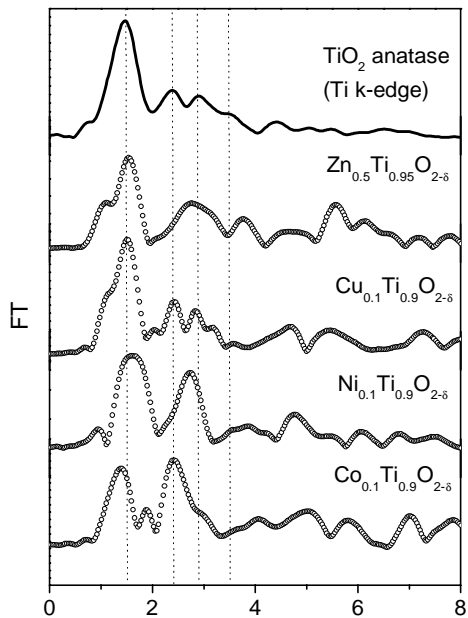

Figure 3: Fourier transforms of doped films and the corresponding to Ti in TiO$_2$ anatase.

**4. Conclusions**

In this work we present an electronic and structural characterization by XAS of TM-doped anatase TiO$_2$ thin films (TM = Co, Ni, Cu, and Zn), ferromagnetic at RT, deposited by PLD. Metallic clusters were neither found in XRD patterns nor in XAS spectra. XANES and EXAFS spectra (performed at the K-edge of each impurity) show that the oxidation state of the impurities is 2+ and the oxygen NN local coordination is similar to that in their respective TM monoxides but, except in the Cu case, the similarity with these oxides only exists in the short range scale. The Fourier transform spectra of the doped films are similar to that of TiO$_2$ anatase suggesting that there is a considerable amount of TM substituting Ti in TiO$_2$ anatase, although the presence of TM-O clusters can not be discarded. Since the TM oxides are paramagnetic at room temperature, the magnetic signals present in the films could be assigned to the MT replacing cations in TiO$_2$ anatase.

Research grants PIPs 6005 and 6032 from CONICET, Red Nacional de Magnetismo (RN3M) and ANPCyT, Argentina, are gratefully acknowledged. This work was dedicated to the memory of Dra. Stella Duhalde.

[1] Y. Matsumoto *et al.*, Science **291**, 854 (2001).
[2] W. K. Park *et al.*, J. Appl. Phys. **91**, 8093 (2002).
[3] See, e.g., H. Weng *et al.*, Phys. Rev. B **73**, 121201 (2006), and refs. therein
[4] S. Duhalde *et al.*, Phys. Rev. B **72**, R161313 (2005).
[5] S. Duhalde et al. J. Phys.: Conference Series (2006), in press.
[6] S.A. Chambers, S.M. Heald and T. Droubay, Phys. Rev. B **67**, 100401(R) (2003).
[7] L. A. Errico, M. Weissmann, and M. Rentería, Phys. Rev. B **72**, 184425 (2005).